\documentclass[12pt]{article}

\usepackage{amsmath,verbatim,amssymb,epsfig,lscape}
\usepackage{graphicx,setspace,float,rotating,longtable}
\usepackage{color}
\usepackage[authoryear]{natbib}
\usepackage[table]{xcolor}
\usepackage{multirow, caption}
\usepackage[colorlinks=true,urlcolor=blue,citecolor=purple,linkcolor=blue,bookmarks=true]{hyperref}

\def\singlespace{\def\baselinestretch{1}\@normalsize}

\begin{document}

\newcommand{\tabincell}[2]{\begin{tabular}{@{}#1@{}}#2\end{tabular}}

\baselineskip=24pt
\begin{center}
{\Large \bf  Optimal Bayesian hierarchical model to accelerate the development of tissue-agnostic drugs and basket trials}
\end{center}

\vspace{2mm}
\begin{center}
{\bf Liyun Jiang$^{1, 2}$, Lei Nie$^{3}$,  Fangrong Yan$^{1}$,  and Ying Yuan$^{2}$}
\end{center}

\begin{center}
$^{1}$Research Center of Biostatistics and Computational Pharmacy, \\China Pharmaceutical University, Nanjing, China\\
$^{2}$Department of Biostatistics, The University of Texas MD Anderson Cancer Center, Houston, TX\\
$^{3}$Center for Drug Evaluation and Research, Food and Drug Administration (FDA), Silver Spring, MD
\vspace{2mm}

\end{center}
\noindent \emph{\textbf{Abstract}}:

Tissue-agnostic trials enroll patients based on their genetic biomarkers, not tumor type, in an attempt to determine if a new drug can successfully treat disease conditions based on biomarkers. The Bayesian hierarchical model (BHM) provides an attractive approach to design phase II tissue-agnostic trials by allowing information borrowing across multiple disease types. In this article, we elucidate two intrinsic and inevitable issues that may limit the use of BHM to tissue-agnostic trials: sensitivity to the prior specification of the shrinkage parameter and the competing ``interest" among disease types in increasing power and controlling type I error. To address these issues, we propose the optimal BHM (OBHM) approach. With OBHM, we first specify a flexible utility function to quantify the tradeoff between type I error and power across disease type based on the study objectives, and then we select the prior of the shrinkage parameter to optimize the utility function of clinical and regulatory interest. OBMH effectively balances type I and II errors, addresses the sensitivity of the prior selection, and reduces the ``unwarranted" subjectivity in the prior selection.  Simulation study shows that the resulting OBHM and its extensions, clustered OBHM (COBHM) and adaptive OBHM (AOBHM), have desirable operating characteristics, outperforming some existing methods with better balanced power and type I error control.  Our method provides a systematic, rigorous way to apply BHM and solve the common problem of blindingly using a non-informative inverse-gamma prior (with a large variance) or priors arbitrarily chosen that may lead to pathological statistical properties.

\vspace{0.5cm}


\section{Introduction}
Tissue-agnostic clinical trials have become increasingly important in the development of targeted therapy and immunotherapy for cancer \citep{Flaherty, Boyiadzis}. Tissue-agnostic trials enroll patients based on their genetic biomarkers (e.g., genetic or molecular aberrations), not tumor type, in an attempt to determine if a new drug can successfully treat cancers based on biomarkers. The US Food and Drug Administration recently granted accelerated approval to Rozlytrek (entrectinib) for treating patients whose cancers have the neurotrophic tyrosine receptor kinase (NTRK) gene fusion, regardless of specific cancer types. The approval embodies a revolutionary paradigm shift in cancer drug development and personalized cancer treatment. The previous tissue-agnostic indications approved by the FDA were pembrolizumab for tumors with microsatellite instability-high (MSI-H) or mismatch repair deficient (dMMR) tumors, and larotrectinib for NTRK gene fusion tumors.  In addition, the National Cancer Institute (NCI) and American Society of Clinical Oncology (ASCO) are sponsoring tissue-agnostic clinical trials in collaboration with industry partners: NCI’s MATCH program and ASCO’s Targeted Agent and Profiling Utilization Registry (TAPUR) study. Each study is enrolling patients based on their genetic biomarkers, not tumor type{\it{s}}. Tissue-agnostic trials are also known as basket trials.

The Bayesian hierarchical model (BHM) provides an attractive approach to design tissue-agnostic clinical trials by allowing information borrowing  across multiple tumor types included in the trial \citep{Thall, Berry}. BHM works by shrinking the tumor-type-specific treatment effect toward the common mean treatment effect, via a shrinkage parameter, thereby borrowing information across tumor types. One {{key}} challenge of using BHM to design tissue-agnostic trials is that it assumes that the investigational drug is tissue agnostic. {{However, this assumption may not be correct,}} i.e., the drug may be effective in some tumor types, but not in other tumor types, despite their sharing the same biomarkers \citep{Flaherty, Tiacci, Prahallad}.  For example, trastuzumab is effective for treating human epidermal growth factor receptor 2 (HER2)-positive breast cancer, but shows little clinical benefit for HER2-positive recurrent endometrial cancer \citep{Fleming} or HER2-positive non-small-cell lung cancer \citep{Gatzemeier}. {{One of the goals for a tissue-agnostic trial is to assess whether the treatment is tissue agnostic in all diseases/tumor types under study or only in some of them.}}

In this paper, we term tumor types that respond favorably to the treatment as ``sensitive", and tumor types that do not respond to the treatment as ``insensitive".  To achieve adaptively information borrowing, a natural approach is to assign the BHM shrinkage parameter a vague prior (with a large variance) and let the trial data determine how much information {{to}} borrow. \citet{Thall} first proposed to use BHM to borrow information across tumor subtypes.  \citet{Berry} applied it to basket trials and suggested to use non-informative prior inverse gamma $IG(0.0005, 0.000005)$. The expectation is that BHM would induce strong shrinkage when the treatment effect is homogenous, but little shrinkage when the treatment effect is heterogeneous across tumor types.  However, \citet{Freidlin} and  \citet{Chu} showed that when there are 10 or fewer tumor types, as is often the case in practice, BHM has difficulty achieving adaptive information borrowing.  Specifically, when a vague prior such as $IG(0.0005, 0.000005)$ is used, BHM tends to always induce strong shrinkage, regardless of if the treatment effect is homogeneous or heterogeneous across tumor types, leading to a substantially inflated type I error or low power when the treatment effect actually is heterogeneous. The reason that BHM has difficulty achieving adaptive information borrowing is that data contain limited information to determine the variation among tumor types (i.e., the shrinkage parameter). This is because the observation to determine the shrinkage parameter is the number of tumor types, not the number of patients. As a result, the operating characteristics of BHM is intrinsically sensitive to and heavily depends on the specification of the prior of the shrinkage parameter.

Thus, the key questions are: what are the desired operation characteristics, and how do we appropriately choose the prior to {{achieve the desired}} operating characteristics? One of the key desired operation characteristics of a design is type I/II error rate control. Given that type I and II errors are competing characteristics, sponsors may choose some desired balance between them in phase II tissue-agnostic trials. In this paper, we propose a utility-based approach to specify the prior of the shrinkage parameter in the BHM, with application to phase II tissue-agnostic trials. A utility function is employed to quantify the tradeoff between type I error and power for each tumor type, defined based on the study objectives. 
The overall mean utility is defined as the weighted average of the tradeoff function across all possible scenarios with homogeneous or heterogeneous treatment effects, where the weight is chosen to reflect the relative importance of each scenario. The optimal prior, and the resulting optimal BHM (OBHM), are obtained by maximizing the mean utility. To achieve better operating characteristics, we further propose clustered OBHM (COBHM), which first groups tumor types into sensitive and insensitive clusters, and then applies BHM within each cluster to borrow information. A similar optimization approach is used to maximize the mean utility.   We also propose adaptive OBHM (AOBHM) that uses data-driven weights in the tradeoff function, based on the Bayesian model averaging (BMA) method,  to reduce the probability of inappropriate information borrowing across tumor types.

To the best of our knowledge, this is the first work that formally considers the tradeoff between type I error and power in this setting, and it provides a systematical and statistically rigorous way to optimize BHM for tissue-agnostic (or basket) trials. {{The major innovation is in}} elucidating the intrinsic sensitivity of BHM to the prior specification of the shrinkage parameter and the inevitable competing ``interest" among tumor types for increasing power and controlling type I errors.  {{Inheriting the advantages of the BMH approach, through the utility function, OBMH systematically and rigorously addresses the previously-noted main issues. It also sets a}} realistic expectation, {{acknowledging the unrealistic expectation that a typical BMH approach could}} borrow information {{across different disease types}}, while strictly controlling type I errors in each tumor type. In addition, OBMH also alleviates the concern of arbitrary or subjective aspects of the prior selection, either in a prespecified or post-hoc fashion. It is useful to reduce and solve the common problem of blindly using a large-variance inverse-gamma prior or a prior arbitrarily specified, which may lead to pathological statistical properties \citep{Gelman}.

{The rest of this paper is organized as follows. In Section 2, we introduce the utility-based approach for OBHM, COBHM and AOBHM. In Section 3, simulation studies, including sensitivity analysis, are carried out to examine the operating characteristics of the proposed approach. A brief discussion is presented in Section 4}.

\section{Methods}
\subsection{BHM}
Consider a phase II tissue-agnostic trial that evaluates the treatment effect of a new antitumor drug in $J$ different tumor types. Let $p_j$ denote the response rate in tumor type $j$. The objective of the trial is to test whether the new drug is effective in each tumor type, with null and alternative hypotheses:
$$
H_0: p_j \leq p_{0,j}\quad versus\quad H_1:p_j \geq p_{1,j}, \qquad j = 1, \cdots, J,
$$
where  $p_{0,j}$ is the null response rate that is deemed futile, and $p_{1,j}$ is the target response rate that is deemed promising for tumor type $j$.

Suppose at an interim analysis time, for tumor type $j$, $n_j$ patients are enrolled, and among them $x_j$ patients responded to the treatment.  BHM can be used to model $p_j$'s as follows,
\begin{eqnarray}\label{bhm}
\begin{aligned}
x_j|p_j \sim Binomial(n_j, p_j),\\
\theta_j  =  \log(\frac{p_j}{1-p_j}) - \log(\frac{p_{0, j}}{1-p_{0, j}}),\\
\theta_j|\theta,\sigma^2  \sim N(\theta, \sigma^2),\\
\theta \sim N(\alpha_0, \tau_0^2),
\end{aligned}
\end{eqnarray}
where $ \alpha_0$, $\tau_0^2$ are hyperparameters, such as  $\alpha_0=0$ and $\tau_0^2=100$. When estimating $p_j$, BHM borrows information across $J$ tumor types by shrinking the tumor-type-specific mean $\theta_j$ toward the common mean $\theta$.  Shrinkage parameter $\sigma^2$ controls the degree of information borrowing. A small value of $\sigma^2$ induces strong information borrowing across tumor types, with the extreme case that $\sigma^2=0$ is identical to directly pooling data across tumor types. A large value of $\sigma^2$ induces little information borrowing, with the extreme case that $\sigma^2 \to \infty$ is identical to an independent analysis approach, which evaluates the treatment in each tumor type independently without any information borrowing.

Borrowing information can be beneficial or detrimental, depending on whether the treatment effect is homogeneous or heterogeneous across tumor types. When the treatment effect $\theta_j$ is homogenous across tumor types (e.g., the treatment is effective for all tumor types), borrowing information significantly improves power. However, when the treatment effect $\theta_j$ is heterogeneous (i.e., the treatment is effective for some tumor types and not effective for other tumor types), borrowing information may substantially inflate the type I error for insensitive tumor types and reduce the power for sensitive tumor types. Thus, it is desirable to borrow information when $\theta_j$ is homogenous, and refrain from borrowing information when $\theta_j$ is heterogeneous. To achieve such adaptive information borrowing,  a natural approach is to assign the shrinkage parameter $\sigma^2$ a vague prior with a large variance, e.g., $IG(0.0005, 0.000005)$ as suggested by \citep{Berry},  and let the trial data determine how much information to borrow, with the intent that BHM induces strong shrinkage when the treatment effect is homogenous, and little shrinkage when the treatment effect is heterogeneous across tumor types (i.e., adaptive information borrowing). Unfortunately, this is not the case when the number of tumor types $J<10$. BHM may still induce strong shrinkage when the treatment effect is heterogeneous, resulting in an inflated type I error for insensitive tumor types and reduced power for sensitive tumor types  \citep{Freidlin, Chu}. Our simulation study shows that BHM with an $IG(0.0005, 0.000005)$ prior can inflate the type I error rate from the nominal level of $10\%$ to over 50\% in some scenarios. The fundamental reason for this issue is that $\sigma^2$ represents the between-tumor-type variance, and thus the observation unit contributing to the estimation of $\sigma^2$ is tumor types, rather than patients. When the number of tumor types $J$ is small, the observed data are insufficient to estimate $\sigma^2$ reliably, even when the number of patients in each tumor type is large. As a result, the operating characteristics of BHM are inevitably sensitive to the prior of $\sigma^2$.  This provides the first motivation for our method.

The second motivation for our method is to recognize the fact that under BHM, the trade-off between type I error and power is also inevitable. As long as we intend to borrow information (for increasing power), the inflation of type I errors is inevitable. This is not only true for the BHM, but for all information-borrowing methods, e.g., power prior \citep{Ibrahim} or commensurate prior \citep{Hobbs}, among others. The reason is simple.  Based on observed data, we cannot, with the probability of 1, correctly determine whether the tumor types are homogeneous or heterogeneous.

Given that the sensitivity to the prior of $\sigma^2$ and the trade-off between type I error and power are both inevitable, a sensible approach is to quantify this trade-off and choose the prior of $\sigma^2$ to optimize the situation, as described in the next section.

\subsection{OBHM}
The prior of $\sigma^2$, denoted as $\pi(\sigma^2)$, has different impacts on sensitive and insensitive tumor types. Let $R$ denote the set of sensitive tumor types, and ${\bar{R}}$ denote the set of insensitive tumor types. When $R$ (or ${\bar{R}}$) is an empty set,  it means that all tumor types are insensitive (or sensitive). Rejecting $H_0$ corresponds to the type I error and power, respectively, for tumor types in ${\bar{R}}$ and $R$. A specification of $\pi(\sigma^2)$ that favors information borrowing may increase power for sensitive tumor types, but at the risk of inflating type I errors for insensitive tumor types. We quantify such power-type-I-error tradeoff using the following utility function:
\begin{equation}\label{uk}
U =\sum_{i\epsilon R}\rho_{i}-\sum_{j\epsilon{\bar{R}}}\{\lambda_{1} \gamma_{j}+\lambda_{2} (\gamma_{j}-\eta) I(\gamma_{j}>\eta)\},
\end{equation}
where $\rho_{i}$ is the power of $i$th (sensitive) tumor type in $R$, $\gamma_{j}$ is the type I error rate of $j$th (insensitive) tumor type in  ${\bar{R}}$, and $\lambda_{1}$ and $\lambda_{2}$ are penalty weights. This utility imposes a penalty of $\lambda_{1}$ for each unit increase of a type I error before it reaches $\eta$, and then imposes a larger penalty of $\lambda_{1}+\lambda_{2}$ after $\gamma_{j}$ exceeds $\eta$. For example, setting $\lambda_{1}=1, \lambda_{2}=2$, and $\eta=0.1$ means that before the type I error of a tumor type in ${\bar{R}}$ reaches 10\%, the penalty of 1\% increase of a type I error is that it cancels out 1\% increase of power of a tumor type in ${{R}}$; after the type I error exceeds 10\%, the penalty of 1\% increase of type I error is that it cancels out 3\% increase of power. This change-point penalty function reflects the common consideration that before a type I error reaches a certain unacceptable high level, it is reasonable to tolerate some type I error inflation in exchange for power gain; after that, type I error inflation is acceptable only when there is a large power gain. If we set $\lambda_2$ to a large value, we essentially control the type I error below $\eta$.

Another desirable property of the proposed utility approach is that it takes into account the ``interest" of all tumor types, reflecting the key feature of the tissue-agnostic trials, and maximizes the overall ``interest." Due to the newness of tissue-agnostic trials, there is no consensus on whether we should uniformly control type I errors for each tumor type at a fixed level.  Our viewpoint is that this should depend on the scenario. For example, in the case that most tumor types are sensitive, it may not be sensible to forbid information borrowing, which sacrifices the power gain of most tumor types, just in order to control the type I error of one or two tumor types. On the other hand, in the case the some tumor types are sensitive and some are insensitive, it may be reasonable to have more strict type I error controls.  In other words, we should take the overall ``interest" into account, and our utility approach reflects such consideration.  

The utility function (\ref{uk}) is not the only way to specify power-type-I-error tradeoff. Other forms of tradeoff function are certainly possible, and they should be customized for the trial. For example, we could factor in the cost effectiveness in the utility, such as
\begin{equation}
U =\sum_{i\epsilon R} Q_i\rho_{i} -\sum_{j\epsilon{\bar{R}}}\{F_{1} \gamma_{j}+F_{2} (\gamma_{j}-\eta) I(\gamma_{j}>\eta)\},
\end{equation}
where $Q_i$ is the potential financial gain for correctly identifying that the drug is effective for $i$th sensitive tumor type for the sponsor and patient care viewpoint, and $F_1$ and $F_2$ are the loss due to incorrectly concluding that the drug is effective for an insensitive tumor type. The loss includes the financial loss for sponsors (e.g., led to the failure of subsequent phase III trials) and wellness loss for in-trial, as well as future, patients (e.g., tolerate side effects and miss potential opportunities to receive other possible treatments). Another more flexible utility function is to allow three regions of tradeoff, e.g., 
\begin{equation}
U =\sum_{i\epsilon R}\rho_{i}-\sum_{j\epsilon{\bar{R}}}\{\lambda_{1} \gamma_{j}+\lambda_{2} (\gamma_{j}-\eta_1) I( \gamma_{j} > \eta_1) +\lambda_{3} (\gamma_{j}-\eta_2) I(\gamma_{j}>\eta_2)\},
\end{equation}
with $\eta_2>\eta_1$, and $\lambda_1, \lambda_2, \lambda_3 >0$. This utility function can be used to represent the case that we have little concern when type I error $\gamma_j \le \eta_1$, some concern when $\eta_1 < \gamma_j \le \eta_2$, and serious concern when $\gamma_j > \eta_2$. When setting $\lambda_2 = \infty$,  we control the type I error for each tumor type at the  level of $\eta_2$.

 As the membership of $R$ and ${\bar{R}}$ is unknown (i.e., {\it a priori}, it is unknown if a tumor type is sensitive or not), $U$ cannot be directly evaluated. Our strategy is to enumerate all possible partitions of $J$ tumor types to form $R$ and ${\bar{R}}$, e.g., $R$ contains $0, 1, \cdots, J$ tumor types, and the remaining tumor types belong to ${\bar{R}}$. The number of possible partitions $G$ depends on whether $J$ tumor types share the same pair of the null response rate and target response rate $(p_{0,j}, p_{1,j})$. In the case that all $J$ tumor types share the same $(p_{0,j}, p_{1,j}) = (p_{0}, p_{1})$, then $G=J+1$; while in the other extreme case that $(p_{0,j}, p_{1,j})$ are distinct across the tumor types, $G=2^J$. For other cases with some tumor types sharing the same $(p_{0,j}, p_{1,j})$, $G$ is between $J+1$ and $2^J$. For example, in our simulation study, we consider $J=4$ tumor types with null response rates $(p_{0,1}, p_{0,2}, p_{0,3},$ $p_{0,4})=(0.05, 0.05, 0.05, 0.15)$ and target response rates $(p_{1,1}, p_{1,2}, p_{1,3},$ $p_{1,4})=(0.20, 0.20, 0.20, 0.30)$. There are total $G=8$ different partitions, as shown in Table \ref{tab:res}. Given a specific partition, $U$ can be evaluated. Let $U_g$ denote the utility corresponding to the $g$th partition, where $g=1, \cdots, G$. We define the mean utility as
\begin{equation}\label{uti}
\bar{U}=\sum_{g=1}^{G}{w_g  U_g},
\end{equation}
where $w_g$ is the weight ascribed to the $g$th partition, with $\sum_{g=1}^{J}{w_g}=1$. The value of $w_g$ should be chosen to reflect the relative importance of each partition. For example, based on preclinical or clinical data, one partition (e.g., only one tumor type is sensitive while all others are insensitive) is believed to be unlikely, then we can assign that partition a low weight.


Our approach is to choose $\pi(\sigma^2)$ to maximize $\bar{U}$. A commonly used conjugate prior for $\sigma^2$ is inverse gamma distribution $IG(a_0, b_0)$. We propose to find $a_0$ and $b_0$ to maximize $\bar{U}$ through grid search. To facilitate this optimization, we re-parameterize the inverse gamma distribution as a scaled inverse-$\chi^2$ distribution, i.e., Inv-$\chi^2(v_0, \sigma_0^2)$, where the scale parameter $\sigma_0^2 = b_0/a_0$ is the prior mean of $\sigma^2$, and the degrees of freedom parameter $v_0 = 2a_0$ can be interpreted as the prior effective sample size. The intuitive interpretation of $v_0$ and $\sigma_0^2$ simplifies choosing the appropriate parameter space for grid search.  As it is generally undesirable to make the prior overly informative, we set the search space for $v_0$ as [0.1, $J$], i.e., the prior effective sample size is from 0.1 to the number of tumor types $J$. We choose 0.1, rather than 0, as the lower boundary to reduce the computational time of the grid search; because the prior effective sample size of 0.1 is already very small, further reducing it has negligible impact on $\bar{U}$ and optimization. To determine the search space for $\sigma_0^2$, we calculate the sample variance of $\theta_j  =  \log(\frac{p_j}{1-p_j}) - \log(\frac{p_{0, j}}{1-p_{0, j}})$, which can be viewed as an empirical Bayes estimate of $\sigma^2$, under each possible partition with $p_j=p_{1, j}$ for sensitive tumor types and $p_j=p_{0, j}$ for insensitive tumor types. The plausible space for $\sigma^2$ is (0, $\hat{\sigma}_{max}^2$], where  $\hat{\sigma}_{max}^2$ is the largest sample variance across all partitions. To allow more flexibility, we set the search space of $\sigma_0^2$ as $(0, 5\hat{\sigma}_{max}^2]$. The specific grid search procedure is shown in the appendix.

\cite{Gelman} noted that blindingly using a vague inverse gamma prior with a very large variance, such as $IG(0.0005, 0.000005)$, in BHM can lead to pathological behavior and recommended using a half-Cauchy prior for $\sigma^2$.
$$
p(\sigma) \propto (1+(\frac{\sigma}{A})^2)^{-1},
$$
where $A$ is the scale parameter. The proposed approach can be directly applied to that prior by choosing $A$ to maximizes the mean utility $\bar{U}$. Our simulation study shows that a half-Cauchy prior yields almost identical operating characteristics as the inverse gamma distribution. Due to its conjugacy and simplicity, inverse gamma may be preferred. Notably, our approach automatically addresses the issue noted by Gelman, because our optimization prevents using an inverse gamma prior with a very large variance.

We now apply the OBHM to design phase II tissue-agnostic trials. Let $N_j$ denote the prespecified maximum sample size for tumor type $j$. A total of $K_j$ interims are planned for tumor type $j$, occurring when its sample size reaches $n_{j,1}, \cdots, n_{j,K_j}$, where $n_{j,K_j}=N_j$. We allow $K_j$ and interim times to vary across tumor types, which is useful when the accrual rate differs across tumor types. Given the interim data $D$, we fit the OBHM and obtain the posterior distribution of $p_j$, and then make the go/no-go decision as follows:
stop the accrual of tumor type $j$ and claim the treatment is not effective for tumor type $j$ if
\begin{equation}\label{stop}
Pr(p_j \le p_{0,j}|~D) > C(n_j),
\end{equation}
where $C(n_j)$ is the probability cutoff calibrated by simulation. Here, we adopt the optimal probability cutoff proposed by the Bayesian optimal phase II (BOP2) design \citep{Zhou},
$$C(n_j)=1-\zeta(\frac{n_j}{N_j})^\delta,$$
where $\zeta$ and $\delta (\ge0)$ are tuning parameters. In the grid search procedure for optimal prior, we use the online BOP2 app available at \url{www.trialdesign.org} to obtain the values of $\delta$ and $\zeta$. After the optimal prior is determined, $\zeta$ is re-calibrated such that under the global null (i.e., all tumor types are insensitive) the type I error for each tumor type is controlled at a prespecified level.

For some trials,  when appropriate, we can also add the superiority stopping rule: stop the accrual of tumor type $j$ and claim the treatment is promising for tumor type $j$, if
\begin{equation}\label{stopsup}
Pr(p_j \ge p_{1,j}|~D) > C_2,
\end{equation}
where $C_2$ is a probability cutoff.

\subsection{COBHM}

By design, OBHM (and also BHM) tends to not borrow information when the treatment effect is heterogeneous. As a result, when one or two tumor types are insensitive, while the majority of the others are sensitive, it prevents information borrowing among sensitive tumor types, which may not be desirable. A more efficient approach is to group tumor types into either sensitive or insensitive homogeneous clusters, and then use OBHM to borrow information within the clusters. This is the idea of clustered OBHM (COBHM), which allows for increasing the power for sensitive tumor types, while controlling the type I error for insensitive tumor types.

At an interim time, based on interim data $D$, we apply the following Bayesian rule to cluster tumor types: a tumor type is allocated to the sensitive cluster ${\cal R}$ if it satisfies
\begin{equation}\label{cluster}
Pr(p_j > \frac{p_{0,j} + p_{1,j}}{2} ~ | ~D) > 0.5\times(\dfrac{n_j}{N_{j}})^\omega,
\end{equation}
otherwise it is allocated to the insensitive cluster $\overline{\cal R}$, where $w>0$ is a tuning parameter. One important feature of this clustering rule is that its probability cutoff is adaptive and depends on the tumor type interim sample size $n_j$. At the early stage of the trial, where $n_j$ is small, we prefer to use a more relaxed (i.e., smaller) cutoff to keep a tumor type in the sensitive cluster to avoid inadvertent stopping, due to sparse data, and to encourage collecting more data on the tumor type. When a trial proceeds, we should use a more strict (i.e., larger) cutoff to avoid incorrectly classifying insensitive tumor types to a sensitive cluster. We recommend set $\omega = 2$ or $3$, which can be further calibrated to fit a specific trial requirement in operating characteristics.
In the above Bayesian clustering rule, the posterior probability $Pr(p_j > \frac{p_{0,j} + p_{1,j}}{2}~|~D)$ is evaluated based on the beta-binomial model,
\begin{eqnarray}
\begin{aligned}
x_j | p_j \sim Binomial(n_j, p_j),  \label{betabinom} \\
p_j \sim Beta(a_1, b_1),
\end{aligned}
\end{eqnarray}
where $a_1$ and $b_1$ are hyperparameters, typically set at a small value (e.g., $a_1=b_1=0.1$) to obtain a vague prior. As a result, the posterior distribution of $p_j$ is given by $Beta(x_j+a_1, n_j-x_j+b_1)$. After clustering, we apply the BHM to each cluster and optimize the prior of $\sigma^2$ by maximizing the mean utility. If ${\cal R}$ or $\overline{\cal R}$ only has one member, we replace BHM with above Beta-Binomial model (\ref{betabinom}).

We focus on the case of clustering the tumor types into two clusters (i.e., sensitive and insensitive clusters) due to the practical consideration that targeted therapy is often either effective (i.e., hit the target) or ineffective (i.e., miss the target). In addition, as the number of tumor types in a trial is typically small ($2-6$), it is not practical to form more than two clusters. Nevertheless, if it is desirable to form more than two clusters, we can use multiple probability cutoffs in (\ref{cluster}) to group the tumor types into multiple clusters (e.g., not, moderately, and highly effective). In addition, other more sophisticated clustering methods, such as K-means, decision tree, and random forest, can also be used to cluster tumor types. We found that the proposed simple Bayesian rule yields comparable or better operating characteristics as these more sophisticated methods. This may be because the data are sparse and the clustering is just an intermediate step of our go/no-go decision making process.

\subsection{AOBHM}
One challenge of OBHM is the requirement of pre-specifying weight $w_g$ for each possible partition for calculating the mean utility. This section presents a data-driven method to automatically determine $w_g$ based on the Bayesian model averaging (BMA) approach \citep{Raftery, Yin}. We referred to the resulting method as adaptive OBHM (AOBHM). The basic idea is to treat each possible partition as a model, and then using BMA to automatically favor the most likely one to optimize the information borrowing. 

Specifically, given a partition $g$, $g=1, \cdots, G$, we employ the procedure in Section 2.2 to determine the OBHM that maximizes the utility under the partition. That is, in  equation (\ref{uti}), we set $w_g=1$ and $w_{g'}=0$, $g'\neq g$, to maximize the utility $\bar{U} =U_g$. Let $M_g$ denote the resulting OBHM. Conditional on the interim data $D$, the posterior probability of $M_g$ is given by
$$
Pr(M_g|D)=\frac{L(D|M_g)Pr(M_g)}{\sum_{i=1}^{G}L(D|M_i)Pr(M_i)},
$$
 where $
Pr(M_g)$ is the prior of $M_g$. In general, we apply the non-informative prior, i.e., $Pr(M_g)=1/G$, when there is no preference for any specific partition. $L(D|M_g)$ is the likelihood of $M_g$, given by 
 $$
L(D|M_g)=\prod_{j=1}^{J} {n_j \choose x_j} p_{j,g}^{x_j}(1-p_{j,g})^{n_j-x_j},
 $$
where $p_{j,g}$ is the response rate for tumor type $j$ given the $g$th partition. 

To make the go/no-go decision for the $j$th indication, the futility stopping rule (\ref{stop}) is calculated as follows:
\begin{equation}
Pr(p_j \le p_{0,j}|~D)=\sum_{g=1}^{G}Pr(p_j \le p_{0,j}|~D, M_g)Pr(M_g|~D).
\end{equation}
As $Pr(p_j \le p_{0,j}|~D, M_g)$ is the decision rule of the OBHM under the $g$th partition, this BMA posterior probability automatically favors the partition with the highest posterior probability through the weight $Pr(M_g|~D)$.  This approach has been previously used by \citep{Yin} to achieve robust dose finding.  

An alternative approach to BMA is the Bayesian model selection: select the $M_g$ with the highest value of $Pr(M_g|~D)$, and then use the OBHM corresponding to that partition to make the go/no-go decision. Given the small sample size, the selection of one partition is associated with large variation, and thus the BMA may be preferred.

\section{Simulation studies}
{\subsection{Simulation setting}}

We evaluated the operating characteristics of the proposed OBHM, COBHM and AOBHM designs through simulation study. We consider $J=4$ tumor types with null response rates $(p_{0,1}, p_{0,2}, p_{0,3},$ $p_{0,4})=(0.05, 0.05, 0.05, 0.15)$ and target response rates $(p_{1,1}, p_{1,2},p_{1,3},p_{1,4})=(0.20, 0.20, 0.20, 0.30)$. We consider one interim analysis and a final analysis, i.e., $K=2$. The maximum sample size for each tumor type is 20, with an interim at the middle.
In utility (\ref{uk}), we set two penalty weights $\lambda_{1}=1, \lambda_{2}=2$, and penalty change point $\eta=0.2$. Given 4 tumor types, there are 8 possible partitions; see Table \ref{tab:res}. For OBHM and COBHM, we set equal weights $w_g=1/8, g=1, \cdots, 8$ for the 8 partitions to define the mean utility (\ref{uti}). Applying the proposed method, the optimal prior for $\sigma^2$ that maximizes the mean utility is $IG(2, 8)$ for the OBHM, and $IG(1, 1.44)$ for the COBHM.  
For AOBHM, we used the noninformative prior and assigned each partition an equal prior probability of 1/8. 
The tuning parameters in the stopping rule  (\ref{stop}) for 4 tumor types are $(\zeta_1, \delta_1)=(0.715, 0.32), (\zeta_2, \delta_2)=(0.715, 0.32), (\zeta_3, \delta_3)=(0.715, 0.32)$ and $(\zeta_4, \delta_4)=(0.7, 0)$ for OBHM, $(\zeta_1, \delta_1)=(0.715, 0.32), (\zeta_2, \delta_2)=(0.715, 0.32), (\zeta_3, \delta_3)=(0.715, 0.32)$ and $(\zeta_4, \delta_4)=(0.72, 0)$ for COBHM, and $(\zeta_1, \delta_1)=(0.73, 0.32), (\zeta_2, \delta_2)=(0.73, 0.32), (\zeta_3, \delta_3)=(0.73, 0.32)$ and $(\zeta_4, \delta_4)=(0.7, 0)$ for AOBHM,  calibrated to control the type I error rate of each tumor type at $10\%$ under the global null. 
We compared proposed designs to an independent design, which evaluates the treatment effect in each tumor type independently (without borrowing information) based on Beta-Binomial model, and the BHM design with $IG(0.0005, 0.000005)$ prior (referred to as BHM) employed by \cite{Berry}.

\subsection{Simulation results}

Table~\ref{tab:res} shows the results of independent, BHM, OBHM, COBHM and AOBHM designs based on 5000 simulations. Scenario 1 is the global null where all tumor types are not sensitive. We calibrated the designs such that each of them controls the type I error rate at 10\% for each tumor type. Scenario 2 is the global alternative that all tumor types are sensitive. BHM, OBHM, COBHM and AOBHM yield substantially higher power than independent design, demonstrating the benefit of borrowing information. BHM has the largest power, however leads to substantial type I error inflation and low power in other scenarios when some tumor types are insensitive, as described below. AOBHM yields higher power than OBHM and COBHM, however, as we will see scenarios 3-8, COBHM has better type I error control. 

In scenario 3, tumor types 1, 2 and 4 are sensitive and tumor type 3 is insensitive. The type I error of BHM for tumor type 3 almost triples those of OBHM, COBHM and AOBHM. 
Compared to OBHM, COBHM has lower type I error and comparable power (i.e., 4.1\% lower, 3.6\% lower and 7\% higher for tumor types 1, 2 and 4). AOBHM has similar performance as OBHM.  In scenario 4, tumor types 1-2 are sensitive and tumor types 3-4 are insensitive. OBHM, COBHM and AOBHM outperform BHM with comparable power for tumor types 1-2 and much lower type I error for tumor types 3 and 4. Remarkably, the type I error of BHM is more than double those of OBHM, COBHM and AOBHM. Compared to OBHM, AOBHM yields higher power and comparable type I error, and COBHM has lower type I error and also slightly lower power.

In scenario 5, tumor types 1 and 4 are sensitive and tumor types 2-3 are insensitive. Although BHM has largest power for tumor types 1 and 4, its type I error for tumor types 2-3 are more than double those of OBHM, COBHM and AOBHM. AOBHM and OBHM have similar performance. 
COBHM shows best performance, demonstrated as higher power (e.g., 0.2\% and 5.58\% higher for  tumor types 1 and 4) and much lower type I error (e.g., 6.92\% and 6.96\% lower for tumor types 2-3) than OBHM. This is because, by separating tumor types into sensitive and insensitive clusters, COBHM is less likely to shrink sensitive and insensitive tumor types together. It thus offers better type I error control.

In scenario 6, tumor types 1-3 are sensitive and tumor type 4 is insensitive. BHM yields largest power for tumor types 1-3, but largely inflates the type I error  for tumor type 4 to 51.22\% which is more than double those of OBHM, COBHM and AOBHM. COBHM has about 2.5\% lower power for tumor types 1-3, but also 4.6\% lower type I error for tumor type 4, than OBHM. AOBHM has slightly higher power for tumor types 1-3 and also higher type I error  for tumor type 4 than OBHM. In scenario 7, tumor type 4 is sensitive and tumor types 1-3 are insensitive. OBHM, COBHM and AOBHM outperform BHM with about 10\%  higher power for tumor type 4 and lower type I error (about 11\% or 15\% lower) for tumor types 1-3. Compared to OBHM and AOBHM, COBHM shows best performance, demonstrated as highest power for tumor type 4 and about 4\% lower type I error for tumor types 1-3. AOBHM is comparable to OBHM, with similar type I error and power. In scenario 8, tumor type 1 is sensitive and tumor types 2-4 are insensitive. Similar to the findings in scenario 7,  OBHM, COBHM and AOBHM outperform BHM with higher power for tumor type 1 and lower type I error for tumor types 2-4. COBHM yields better performance than OBHM and AOBHM, demonstrated as higher power and lower type I error. 


\subsection{Sensitivity analysis}

We studied the sensitivity of proposed OBHM, COBHM and AOBHM designs with respect to (i) penalty weights $(\lambda_{1}, \lambda_{2})$, (ii) the penalty change point $\eta$ in utility (\ref{uk}), and (iii) values of scenario weights $w_g$ in average utility (\ref{uti}).

Table \ref{tab:sens1} shows the simulation results when $(\lambda_{1}, \lambda_{2})=(1, 1)$, which are generally similar to those shown in Table \ref{tab:res} when $(\lambda_{1}, \lambda_{2})=(1, 2)$. That is, COBHM outperforms OBHM with a lower type I error rate, and comparable or higher power. AOBHM yields higher power than OBHM when all tumor types are sensitive, and comparable performance in other scenarios.

Table \ref{tab:sens2} shows the simulation results when $\eta=0.15$. OBHM, COBHM and AOBHM are generally robust, in particular the latter two designs. We note slightly more variation for OBHM. For example, compared to the results with $\eta=0.2$ (i.e., Table \ref{tab:res}), the type I error of OBHM is smaller (e.g, scenarios 4, 6, 7), due to a higher penalty for type I error inflation when it exceeds 0.15, but as a tradeoff, the power is also lower. 

Table \ref{tab:sens3} shows the simulation results with a large weight $w=0.85$ assigned to global alternative that all tumor types are sensitive. This weight represents a strong prior belief that the drug is tissue-agnostic and effective for all tumor types. We can see that the results now strongly favor scenario 2, where all tumor types are sensitive.  Compared to the results in Table \ref{tab:res}, the power of OBHM are increased by about 10\% and 20\% for tumor types 1-3 and type 4, respectively, and the power of COBHM are improved by 3.46\% for tumor type 4. As expected, the type I error rates are inflated in scenarios 3-8, especially for OBHM. As these scenarios are regarded as unlikely, they may be of limited concern.

We also applied the proposed approach to optimize the half-Cauchy prior. The simulation results are shown in Table \ref{tab:A1} in Appendix, which are similar to those using inverse gamma prior. These results show that the proposed approach is not sensitive to the function form of the prior, as the prior is calibrated to maximize the same utility. In other words, the specification of utility often matters more than the function form of the prior.  We also consider the trials with 3 tumor types. The results are provided in Table \ref{tab:A2} in the appendix, showing that OBHM, COBHM and AOBHM have desirable operating characteristics similar to Table \ref{tab:res}.

\section{Discussion}
We developed a utility-based approach to optimize BHM and applied it to a phase II tissue-agnostic trials. We proposed a utility function to account for the power-type-I-error tradeoff across tumor types, reflecting the distinct features and interplays between tumor types in a tissue-agnostic trial. The utility is flexible and can be customized to reflect specific trial considerations. The proposed OBHM, COBHM and AOBHM provide statistically rigorous and principled approaches to address the intrinsic power-type-I-error tradeoff of information borrowing. They are useful to reduce and solve the common problem of blindingly using a non-informative inverse-gamma prior (with a large value) or arbitrarily choosing a prior in the BHM. The simulation study shows that OBHM, COBHM and AOBHM have better operating characteristics than the BHM (using a vague inverse gamma prior). Among three proposed methods, we recommend COBHM because it offers better balance in power gain and type I error control, owing to its clustering-and-then-borrowing strategy to separate sensitive tumor types from insensitive tumor types. Compared to OBHM, AOBHM yields power gain when all tumor types are sensitive, otherwise is generally comparable. Note that because of the power-type-I-error tradeoff, it is impossible to have one method dominate the others in both type I error control and power. We applied our method to design phase II tissue-agnostic trials, but it can be used in other applications where the BHM is appropriate. In addition, the proposed design can be extended for ordinal, continual, and survival endpoints, which is the topic of our future research.



\bigskip
\bigskip
\noindent{ \bf Disclaimer} \\
This article reflects the views of the author, and it should not be construed to represent FDA views or policies.

\clearpage
\begin{center}
\renewcommand\arraystretch{1.25}
\setlength{\tabcolsep}{17pt}
\begin{longtable}{cccccc}
\caption{Simulation results of Independent, BHM, OBHM, COBHM and AOBHM designs in eight scenarios with null response rates $(p_{0,1}, p_{0,2}, p_{0,3}, p_{0,4})=(0.05, 0.05, 0.05, 0.15)$ and target response rates $(p_{1,1}, p_{1,2},p_{1,3},p_{1,4})=(0.20, 0.20, 0.20, 0.30)$. 
Sensitive tumor types are in bold.} \label{tab:res}\\
\hline
 \multirow{2}{*}{{Scenario}} & \multirow{2}{*}{{Methods}} & \multicolumn{4}{c}{\tabincell{c}{Probability (\%) of claiming \\  treatment effective}}\\ \cline{3-6}
& & {1} & { 2} & {3} & { 4}  \\ \hline
1($H_0$)&&0.05 &0.05 &0.05 &0.15 \\
 \multirow{6}{*}{} &Independent &9.02 	&9.46	&9.42	&10.98\\
 & BHM &10.52 	&10.72	 &10.66	&10.32\\
& OBHM &10.28 	&10.04	&10.36	&10.44\\
& COBHM &10.10  	&9.84	&10.02	&10.86\\
& AOBHM & 10.32 	&10.16	&10.54	&10.56\\
& & & & & \\
2($H_1$)&&\textbf{0.20} &\textbf{0.20} &\textbf{0.20} &\textbf{0.30} \\
\multirow{6}{*}{}& Independent & 62.44 	&62.42	&63.52	&58.72\\
 & BHM &94.80 	&95.24	&95.36	&88.48\\
& OBHM &85.52	&86.28	&86.12	&68.88\\
& COBHM &83.94 	&84.90	&84.76	&71.08\\
&AOBHM &88.82 	&89.28	&86.92	&74.96\\
& & & & & \\
3&&\textbf{0.20} &\textbf{0.20} &0.05 &\textbf{0.30} \\
\multirow{6}{*}{}& Independent &62.44 	&62.42	&9.42	&58.72\\
 & BHM &89.74 	&89.74	&64.80	&80.32\\
& OBHM &83.76 	&84.62	&21.42	&61.86\\
& COBHM &79.66 	&81.02	&18.54	&68.86\\
&AOBHM&84.56 	&85.36	&23.62	&65.10\\
& & & & & \\
4& &\textbf{0.20} &\textbf{0.20} &0.05 &0.15\\
\multirow{6}{*}{} & Independent &62.44 	&62.42	&9.42	&10.98\\
 & BHM &80.48 	&81.12	&41.36	&36.68\\
& OBHM &80.96	& 81.30	& 20.74	&15.40\\
& COBHM &78.54 	&79.48	&16.26	&14.74\\
&AOBHM &82.10 	&82.68	&20.98	&17.36\\
& & & & & \\
5&&\textbf{0.20} &0.05 &0.05 &\textbf{0.30}\\
\multirow{6}{*}{} &Independent &62.44 	&9.46	&9.42	&58.72\\
 & BHM &79.18 	&44.52	&44.42	 &67.26\\
& OBHM &77.10 	&19.40	&19.94	&58.90\\
& COBHM &77.30 	&12.48	&12.98	&64.48\\
&AOBHM&77.66 	&20.48	&20.20	&60.18\\
& & & & & \\
6& &\textbf{0.20} &\textbf{0.20} &\textbf{0.20} &0.15\\
\multirow{6}{*}{} & Independent &62.44 	&62.42	&63.52	&10.98\\
 & BHM &89.04 	&89.48	&89.68	&51.22\\
& OBHM &84.62 	&85.28	&85.12	&20.34\\
& COBHM &82.18 	&82.96	&82.86	&15.74\\
&AOBHM &85.56 	&86.28	&85.74	&24.10\\
& & & & & \\
7&&0.05 &0.05 &0.05 &\textbf{0.30}\\
 \multirow{6}{*}{} &Independent &9.02 	&9.46	&9.42	&58.72\\
 & BHM &26.64 	&26.50	&26.34	&47.90\\
& OBHM &15.08 	 &14.46	&15.52	&57.70\\
& COBHM	 &11.16 	&10.58	&10.90	&59.02\\
& AOBHM &15.20 	&14.56	&15.32	&57.86\\
& & & & & \\
8 &&\textbf{0.20} &0.05 &0.05 &0.15 \\
 \multirow{6}{*}{} &Independent &62.44 	&9.46	&9.42	&10.98\\
 & BHM &63.62 	&25.34	&25.50	&23.10\\
& OBHM &69.84 	&17.22	&17.86	&11.58\\
& COBHM &75.70 	&11.68	&11.58	&13.18\\
& AOBHM &71.64 	&18.02	&18.34	&13.00\\
\hline
\end{longtable}
\end{center}

\clearpage
\begin{center}
\renewcommand\arraystretch{1.25}
\setlength{\tabcolsep}{17pt}
\begin{longtable}{cccccc}
\caption{Sensitivity analysis of OBHM, COBHM and AOBHM designs based on different penalty weights $\lambda_{1}=1$ and  $\lambda_{2}=1$.} \label{tab:sens1}\\
\hline
 \multirow{2}{*}{{Scenario}} & \multirow{2}{*}{{Methods}} & \multicolumn{4}{c}{\tabincell{c}{Probability (\%) of claiming \\ treatment effective}}\\ \cline{3-6}
& & {1} & { 2} & {3} & { 4} \\ \hline
1($H_0$)&&0.05 &0.05 &0.05 &0.15 \\
 \multirow{6}{*}{}
& OBHM &10.24  	&9.72	&10.02	&10.32\\
& COBHM &10.10  	&9.84	&10.02	&10.86\\
& AOBHM &10.36 	&10.10	&10.50	&10.72\\
& & & & & \\
2($H_1$)&&\textbf{0.20} &\textbf{0.20} &\textbf{0.20} &\textbf{0.30} \\
\multirow{6}{*}{}
& OBHM &85.52 	&86.28	&86.20	&60.64\\
& COBHM &83.94 	&84.90	&84.76	&71.08\\
&AOBHM &90.02 	&89.98	&87.00	&75.60\\
& & & & & \\
3&&\textbf{0.20} &\textbf{0.20} &0.05 &\textbf{0.30} \\
\multirow{6}{*}{}
& OBHM &83.68 	&84.68	&21.38	&58.94\\
& COBHM &79.66 	&81.02	&18.54	&68.86\\
&AOBHM&84.92 	&85.42	&24.10	&65.56\\
& & & & & \\
4& &\textbf{0.20} &\textbf{0.20} &0.05 &0.15\\
\multirow{6}{*}{} 
& OBHM &80.70 	&81.62	&20.64	&12.10\\
& COBHM &78.54 	&79.48	&16.26	&14.74\\
&AOBHM&82.22 	&82.66	&21.00	&17.66\\
& & & & & \\
5&&\textbf{0.20} &0.05 &0.05 &\textbf{0.30}\\
\multirow{6}{*}{} 
& OBHM &75.82 	&19.12	&19.68	&57.90\\
& COBHM &77.30 	&12.48	&12.98	&64.48\\
&AOBHM&77.74 	&20.68	&20.24	&60.84\\
& & & & & \\
6& &\textbf{0.20} &\textbf{0.20} &\textbf{0.20} &0.15\\
\multirow{6}{*}{} 
& OBHM &84.70 	&85.34	&85.30	&14.32\\
& COBHM &82.18 	&82.96	&82.86	&15.74\\
&AOBHM &85.92 	&86.54	&85.72	&24.60\\
& & & & & \\
7&&0.05 &0.05 &0.05 &\textbf{0.30}\\
 \multirow{6}{*}{} 
& OBHM &13.62 	&13.20	&14.08	&57.62\\
& COBHM &11.16 	&10.58	&10.90	&59.02\\
& AOBHM &15.36 	&14.58	&15.44	&58.06\\
& & & & & \\
8 &&\textbf{0.20} &0.05 &0.05 &0.15 \\
 \multirow{6}{*}{} 
& OBHM &70.94 	&16.86	&17.46	&10.64\\
& COBHM &75.70 	&11.68	&11.58	&13.18\\
& AOBHM &71.60 	&18.06	&18.32	&13.52\\
\hline
\end{longtable}
\end{center}


\clearpage
\begin{center}
\renewcommand\arraystretch{1.25}
\setlength{\tabcolsep}{17pt}
\begin{longtable}{cccccc}
\caption{Sensitivity analysis of OBHM, COBHM and AOBHM designs based on a more stringent penalty change point $\eta=0.15$.} \label{tab:sens2}\\
\hline
 \multirow{2}{*}{{Scenario}} & \multirow{2}{*}{{Methods}} & \multicolumn{4}{c}{\tabincell{c}{Probability (\%) of claiming \\ treatment effective}}\\ \cline{3-6}
& & {1} & { 2} & {3} & { 4}  \\ \hline
1($H_0$)&&0.05 &0.05 &0.05 &0.15 \\
 \multirow{6}{*}{}
& OBHM &10.24  	&9.72	&10.02	&10.32\\
& COBHM &10.10  	&9.88  	&9.98	&10.64\\
& AOBHM &10.16  	&9.92  	&9.86	&10.56\\
& & & & & \\
2($H_1$)&&\textbf{0.20} &\textbf{0.20} &\textbf{0.20} &\textbf{0.30} \\
\multirow{6}{*}{}
& OBHM &85.52 	&86.28	&86.20	&60.64\\
& COBHM &83.98 	&84.88	&84.68	&70.22\\
&AOBHM &88.56 	&88.18	&86.50	&73.86\\
& & & & & \\
3&&\textbf{0.20} &\textbf{0.20} &0.05 &\textbf{0.30} \\
\multirow{6}{*}{}
& OBHM &83.68 	&84.68	&21.38	&58.94\\
& COBHM &79.56 	&81.04	&18.56	&67.64\\
&AOBHM&84.40 	&85.10	&22.50	&64.08\\
& & & & & \\
4& &\textbf{0.20} &\textbf{0.20} &0.05 &0.15\\
\multirow{6}{*}{} 
& OBHM &80.70 	&81.62	&20.64	&12.10\\
& COBHM &78.58 	&79.52	&16.28	&14.06\\
&AOBHM&81.58 	&82.44	&20.64	&16.58\\
& & & & & \\
5&&\textbf{0.20} &0.05 &0.05 &\textbf{0.30}\\
\multirow{6}{*}{} 
& OBHM &75.82 	&19.12	&19.68	&57.90\\
& COBHM &77.28 	&12.38	&12.90	&63.48\\
&AOBHM&76.82 	&19.34	&19.72	&58.98\\
& & & & & \\
6& &\textbf{0.20} &\textbf{0.20} &\textbf{0.20} &0.15\\
\multirow{6}{*}{} 
& OBHM &84.70 	&85.34	&85.30	&14.32\\
& COBHM &82.22 	&82.98	&82.86	&15.14\\
&AOBHM &85.36 	&85.96	&85.36	&23.50\\
& & & & & \\
7&&0.05 &0.05 &0.05 &\textbf{0.30}\\
 \multirow{6}{*}{} 
& OBHM &13.62 	&13.20	&14.08	&57.62\\
& COBHM &11.04 	&10.54	&10.86	&58.54\\
& AOBHM &13.86 	&13.52	&13.74	&57.78\\
& & & & & \\
8 &&\textbf{0.20} &0.05 &0.05 &0.15 \\
 \multirow{6}{*}{} 
& OBHM &70.94 	&16.86	&17.46	&10.64\\
& COBHM &75.78 	&11.62	&11.72	&12.82\\
& AOBHM &71.48 	&17.52	&17.74	&11.32\\
\hline
\end{longtable}
\end{center}


\clearpage
\begin{center}
\renewcommand\arraystretch{1.25}
\setlength{\tabcolsep}{17pt}
\begin{longtable}{cccccc}
\caption{Sensitivity analysis of OBHM and COBHM designs based on a large weight $w=0.85$ assigned to global alternative that all tumor types are sensitive (i.e., scenario 2).} \label{tab:sens3}\\
\hline
 \multirow{2}{*}{{Scenario}} & \multirow{2}{*}{{Methods}} & \multicolumn{4}{c}{\tabincell{c}{Probability (\%) of claiming \\ treatment effective}}\\ \cline{3-6}
& & {1} & { 2} & {3} & { 4} \\ \hline
1($H_0$)&&0.05 &0.05 &0.15 &0.15 \\
\multirow{2}{*}{}
& OBHM &10.26  	&9.90  	&9.94  	&9.82\\
& COBHM &9.96 	&10.00  	&9.78	&10.16\\
& & & & & \\
2($H_1$)&&\textbf{0.20} &\textbf{0.20} &\textbf{0.20} &\textbf{0.30} \\
\multirow{6}{*}{}
& OBHM &96.08 	&95.78	&96.14	&88.42\\
& COBHM &83.50 	&84.30	&84.34	&74.54\\
& & & & & \\
3&&\textbf{0.20} &\textbf{0.20} &0.05 &\textbf{0.30} \\
\multirow{6}{*}{}
& OBHM &90.94 	&90.92	&64.56	&81.06\\
& COBHM &79.04 	&80.16	&18.84	&73.08\\
& & & & & \\
4& &\textbf{0.20} &\textbf{0.20} &0.05 &0.15\\
\multirow{6}{*}{} 
& OBHM &82.98 	&83.04	&44.36	&35.68\\
& COBHM &77.74 	&78.66	&13.92	&16.84\\
& & & & & \\
5&&\textbf{0.20} &0.05 &0.05 &\textbf{0.30}\\
\multirow{6}{*}{} 
& OBHM &80.98 	&42.26	&42.52	&68.94\\
& COBHM &77.00 	&13.22	&13.14	&68.36\\
& & & & & \\
6& &\textbf{0.20} &\textbf{0.20} &\textbf{0.20} &0.15\\
\multirow{6}{*}{} 
& OBHM &90.98 	&91.32	&91.26	&50.52\\
& COBHM &79.52 	&80.42	&79.98	&17.66\\
& & & & & \\
7&&0.05 &0.05 &0.05 &\textbf{0.30}\\
 \multirow{6}{*}{} 
& OBHM &22.44 	&21.86	&22.18	&50.70\\
& COBHM &11.48 	&11.06	&11.30	&57.52\\
& & & & & \\
8 &&\textbf{0.20} &0.05 &0.05 &0.15 \\
 \multirow{6}{*}{} 
& OBHM &67.68 	&24.84	&24.72	&22.40\\
& COBHM &74.84 	&11.90	&11.52	&14.78\\
\hline
\end{longtable}
\end{center}


\clearpage
\begin{center}
\noindent \textbf{Appendix}
\end{center}

\noindent{\textit{A. grid search for optimal prior}\\
We re-parameterize the inverse gamma distribution $IG(a_0, b_0)$ by a scaled inverse-$\chi^2$ distribution, i.e., $\sigma^2 \sim$ Inv-$\chi^2(v_0, \sigma_0^2)$, where degrees of freedom $v_0 = 2a_0$ and scale $\sigma_0^2 = b_0/a_0$. We identify optimal $(\sigma_0, v_0)$ through a grid search, and then obtain optimal $(a_0, b_0)$ by equations $a_0=v_0/2$ and $b_0=v_0\cdot\sigma_0^2/2$. As it is generally undesirable to make the prior overly informative, we set the search space for $v_0$ as [0.1, $J$], i.e., the prior effective sample size is from 0.1 to the number of tumor type $J$. To determine the search space for $\sigma_0^2$, we calculate the sample variance of $\theta_j  =  \log(\frac{p_j}{1-p_j}) - \log(\frac{p_{0, j}}{1-p_{0, j}})$, which can be viewed as an empirical Bayes estimate of $\sigma^2$, under each possible partition with $p_j=p_{1, j}$ for sensitive tumor types and $p_j=p_{0, j}$ for insensitive tumor types. The plausible space for $\sigma^2$ is (0, $\hat{\sigma}_{max}^2$], where  $\hat{\sigma}_{max}^2$ is the largest sample variance across all partitions. To allow more flexibility, we set the search space of $\sigma_0^2$ as $(0, 5\hat{\sigma}_{max}^2]$. Suppose the search ranges for $\sigma_0$ and $v_0$ are $(\sigma_0^{(1)}, \cdots, \sigma_0^{(L)})$ and $(v_0^{(1)}, \cdots, v_0^{(M)})$, respectively. The optimal $(a_0, b_0)$ that maximizes the mean utility $\bar{U}$ is obtained based on the following procedure:
\begin{enumerate}
\item Given a specific grid $(\sigma_0^{(l)}, v_0^{(m)})$, compute the power and type I error rates under partition $g$, and obtain the utility $U_g(\sigma_0^{(l)}, v_0^{(m)})$ based on equation (\ref{uk}).
\item After utilities under all partitions are obtained, compute the mean utility $\bar{U}(\sigma_0^{(l)}, v_0^{(m)})$ based on equation (\ref{uti}).
\item Identify the $(\sigma_0^*, v_0^*)$ that yields the largest mean utility $\bar{U}$.
\item Obtain optimal $a_0$ and $b_0$, based on equations $a_0=v_0^*/2$ and $b_0=v_0^*\cdot \sigma_0^{*2}/2$.
\end{enumerate}}

The grid search procedure for the half-Cauchy prior is similar to that for the inverse gamma prior. The main difference is that we only need to search a one-dimensional grid for scale parameter $A$.

\clearpage

\setcounter{table}{0}
\renewcommand{\thetable}{A\arabic{table}}

\begin{center}
\renewcommand\arraystretch{1.25}
\setlength{\tabcolsep}{17pt}
\begin{longtable}{cccccc}
\caption{Simulation results of OBHM, COBHM, and AOBHM with half-Cauchy prior under $\eta=0.20$, $\lambda_{1}=1, \lambda_{2}=2$, and $w_g=1/8, g=1, \cdots, 8$.} \label{tab:A1}\\
\hline
 \multirow{2}{*}{{Scenario}} & \multirow{2}{*}{{Methods}} & \multicolumn{4}{c}{\tabincell{c}{Probability (\%) of claiming \\ treatment effective}} \\ \cline{3-6}
& & {1} & { 2} & {3} & { 4} \\ \hline
1($H_0$)&&0.05 &0.05 &0.05 &0.15 \\
 \multirow{6}{*}{}
& OBHM &9.98  	&9.86  	&9.94	&10.80\\
& COBHM &9.90  	&9.84  	&9.78	&10.76\\
& AOBHM &10.20  	&9.92  	&9.82	&10.82\\
& & & & & \\
2($H_1$)&&\textbf{0.20} &\textbf{0.20} &\textbf{0.20} &\textbf{0.30} \\
\multirow{6}{*}{}
& OBHM &86.82 	&87.60	&87.62	&73.70\\
& COBHM &83.92 	&84.86	&84.70	&69.94\\
&AOBHM &91.16 	&90.78	&87.76	&75.84\\
& & & & & \\
3&&\textbf{0.20} &\textbf{0.20} &0.05 &\textbf{0.30} \\
\multirow{6}{*}{}
& OBHM &84.54 	&85.34	&29.36	&64.30\\
& COBHM &79.52 	&81.02	&18.46	&66.36\\
&AOBHM&85.34 	&85.86	&30.46	&65.18\\
& & & & & \\
4& &\textbf{0.20} &\textbf{0.20} &0.05 &0.15\\
\multirow{6}{*}{} 
& OBHM &81.74 	&81.70	&22.56	&15.54\\
& COBHM &78.48 	&79.42	&15.96	&13.56\\
&AOBHM&81.88 	&81.86	&22.94	&16.60\\
& & & & & \\
5&&\textbf{0.20} &0.05 &0.05 &\textbf{0.30}\\
\multirow{6}{*}{} 
& OBHM &79.40 	&21.78	&21.98	&60.50\\
& COBHM &77.24 	&12.44	&12.90	&61.26\\
&AOBHM&79.64 	&22.82	&22.22	&60.68\\
& & & & & \\
6& &\textbf{0.20} &\textbf{0.20} &\textbf{0.20} &0.15\\
\multirow{6}{*}{} 
& OBHM &84.82 	&85.44	&85.24	&22.66\\
& COBHM &82.02 	&82.90	&82.72	&15.22\\
&AOBHM &86.32 	&86.62	&85.30	&25.38\\
& & & & & \\
7&&0.05 &0.05 &0.05 &\textbf{0.30}\\
 \multirow{6}{*}{} 
& OBHM &17.24 	&16.48	&17.38	&58.40\\
& COBHM &11.04 	&10.62	&10.90	&58.56\\
& AOBHM &17.50 	&16.86	&17.52	&58.44\\
& & & & & \\
8 &&\textbf{0.20} &0.05 &0.05 &0.15 \\
 \multirow{6}{*}{} 
& OBHM &70.14	&18.28	&18.22	&12.32\\
& COBHM &75.66 	&11.60	&11.54	&11.80\\
& AOBHM &70.16 	&18.72	&18.48	&12.38\\
\hline
\end{longtable}
\end{center}


\clearpage
\begin{center}
\renewcommand\arraystretch{1.15}
\setlength{\tabcolsep}{17pt}
\begin{longtable}{ccccc}
\caption{Simulation results with three tumor types for Independent, BHM, OBHM, COBHM and AOBHM designs in four scenarios with null response rates $p_{0,1}=p_{0,2}=p_{0,3}=0.05$ and target response rates $p_{1,1}=p_{1,2}=p_{1,3}=0.20$. $\eta=0.20$, $\lambda_{1}=1, \lambda_{2}=2$, and equal scenario weights $w_1=w_2=w_3=w_4=0.25$. Sensitive tumor types are in bold.} \label{tab:A2}\\
\hline
 \multirow{2}{*}{{Scenario}} & \multirow{2}{*}{{Methods}} & \multicolumn{3}{c}{\tabincell{c}{Probability (\%) of claiming \\ treatment effective}} \\ \cline{3-5}
& & {1} & { 2} & {3} \\ \hline
1($H_0$) & &0.05 &0.05 &0.05\\
\multirow{4}{*}{}
& Independent &9.02 	&9.46	&9.42	\\
& BHM &10.26 	&10.44	&10.14	\\
& OBHM &9.94	&9.84	&10.00\\
& COBHM & 10.12  	&9.86  	&9.62\\
&AOBHM	&10.78 	&10.40	&10.58\\
& & & & \\
2($H_1$)& &\textbf{0.20} &\textbf{0.20} &\textbf{0.20}\\
\multirow{4}{*}{}
& Independent &62.44 	&62.42	&63.52\\
& BHM &93.24 	&93.40	&93.72	\\
& OBHM &85.36	&86.12	&86.04\\
& COBHM &83.82 	&84.66	&84.58\\
&AOBHM	&90.86 	&90.10	&86.10\\
& & & & \\
3& &\textbf{0.20} &\textbf{0.20} &0.05\\
\multirow{4}{*}{}
& Independent &62.44 	&62.42	&9.42\\
& BHM &83.86 	&83.92	&50.62\\
& OBHM & 83.08	&83.52	&21.18\\
& COBHM &77.96 	&79.28	&17.78\\
&AOBHM	&83.88 	&83.96	&21.10\\
& & & & \\
4& &\textbf{0.20} &0.05 &0.05\\
\multirow{4}{*}{}
& Independent &62.44 	&9.46	&9.42\\
& BHM &65.90 	&28.14	&27.76	\\
& OBHM &71.94	&18.00	&18.52\\
& COBHM & 77.54 	&10.72	&10.50	\\
&AOBHM	&72.58 	&19.00	&19.10 \\
\hline
\end{longtable}
\end{center}

\end{document}